\documentclass[twocolumn,letteredappendix%,linenumbers
]{aastex631}

\usepackage{amsmath}

\shorttitle{FRB repeaters go quiet within a few years}
\shortauthors{Yamasaki}

\begin{document}

\title{A Butterfly Collector's View of Fast Radio Bursts:
Repeating Sources Go Quiet within a Few Years}

\author[0000-0002-1688-8708]{Shotaro Yamasaki}
\affiliation{Department of Physics, National Chung Hsing University, Taichung, Taiwan}

\correspondingauthor{Shotaro Yamasaki}
\email{shotaro.s.yamasaki@gmail.com}

\begin{abstract}
How long repeating fast radio burst (FRB) sources stay active is key to their origins but poorly known as a population. We test this with unseen species estimation, a method first used to predict how many new butterfly species a further collecting trip would find. If each source keeps the same mean rate over time, the number of sources seen $k+1$ times in one window predicts how many bursts the sources seen $k$ times will emit in a later window. It assumes neither the burst rate distribution nor source distances. For the five years of CHIME/FRB Catalog~2, we measure the ratio $R$ of observed to predicted bursts against the gap $\Delta$ between windows. A steady mean rate gives $R = 1$ at every $\Delta$, so a change of $R$ with $\Delta$ means that the sources change. $R$ is close to unity for $\Delta \lesssim 60$ days and drops by half within one year. Neither steady nor randomly fluctuating sources reproduce this decline, unless the latter are mostly silent. Three models explain the data equally well: a lifetime of about three years, on/off activity with two-year active phases, or a mix of short- and long-lived sources. Fitted to the first three years alone, they correctly predict the further decline of $R$ in the last two years. Longer monitoring of $R$, and of whether silent sources come back to life, will distinguish them. Unlike butterfly species, FRB sources do not wait for the next collecting trip, so follow-up should start soon.
\end{abstract}


\section{Introduction} \label{sec:intro}
Fast radio bursts (FRBs) are millisecond radio flashes from extragalactic sources \citep{Lorimer2007,CordesChatterjee2019,Petroff2022,Zhang2023}. A small fraction of the sources repeat, while most have been seen only once \citep{CHIME2021cat1,CHIME2026cat2}. Repeaters and one-off sources differ in burst morphology \citep{Pleunis2021}. Whether all sources repeat is still debated. Some studies find that a single population of repeaters can explain the data \citep{Ravi2019,James2023,Yamasaki2024,BeniaminiKumar2025,Cook2026,Hoffmann2026,Yamasaki2026sc}, while others find evidence for sources that never repeat \citep{Dong2026,Sirota2026}.
 
Individual repeaters show that activity is not constant. FRB~20180916B and FRB~20121102A are active in cycles of 16.3 and about 160 days \citep{CHIME2020periodic,Rajwade2020}. FRB~20121102A and FRB~20201124A had short periods of very high activity separated by quiet phases \citep{Li2021,Lanman2022,Xu2022}. However, these examples come from a few well-studied repeaters that are nearby or unusually active. How the activity of the repeater population as a whole changes with time is poorly known. 
Young neutron stars, in particular magnetars, have been proposed as the engines of FRBs \citep[see][for a review]{Zhang2023}, while observations and theoretical models suggest that magnetar bursting activity varies over time, with outbursts that fade over months to years \citep{KaspiBeloborodov2017,CotiZelati2018,Keskin2025}. Measuring the population-wide evolution of repeater activity can therefore constrain the magnetar scenario and determine whether apparently one-off sources could be repeaters that have become inactive.
 
CHIME/FRB is a transit telescope that observes the whole northern sky every day. Its second catalog contains 4539 bursts from 3641 sources, of which 83 are repeaters, all found in one uniform survey over five years \citep{CHIME2026cat2}. This makes population studies possible \citep[e.g.,][]{Cook2026,Sun2026morph,Dong2026,ZhangZhang2026,Sirota2026,Yamasaki2026sc}.

Unseen species estimation is a method of statistics that predicts what a later sample will contain from the counts in an earlier sample, even when that sample is far from complete \citep{Fisher1943,Good1953,Good1956,Robbins1956}. \citet{Fisher1943} studied the butterflies that Corbet collected in Malaya over two years. From the numbers of species caught once, twice, and so on, they asked how many new species a further collecting trip would find. Later, \citet{Efron1976} estimated how many words Shakespeare knew but never used, and \citet{Thisted1987} predicted the number of new words in a newly found poem to test whether Shakespeare wrote it. The central identity of the method is the Robbins formula \citep{Robbins1956}. If each species is caught at a constant rate, the number of species caught $k+1$ times predicts how many individuals the species caught $k$ times will yield in a later sample, for any distribution of rates.

In astronomy, \citet{Ambartsumian1970} used the numbers of stars seen to flare once and twice to estimate how many flare stars in the Pleiades had not yet flared. \citet{Kiang1987} pointed out that this is the same problem as counting the words Shakespeare never used. \citet{Mnatsakanian1988} extended this approach to predict the future flare activity of star clusters. These studies assume that each source keeps a constant rate, just as a butterfly species is always there to be caught.

An FRB survey works in the same way as a butterfly collection. The telescope collects bursts, and each source plays the role of a species. Unlike a butterfly species, however, the activity of an FRB source may change with time. Here we turn the assumption of a constant long-term rate into the quantity we measure. Bursts of well-studied repeaters are known to cluster on timescales of subseconds to hours \citep[e.g.,][]{Oppermann2018,Cruces2021,TotaniTsuzuki2023}, and so are those of CHIME repeaters \citep{Yamasaki2026sc,McKinven2026}. We ask instead whether their mean rate, averaged over days, stays the same over months to years. We predict future bursts of FRB sources from past detections and ask how the prediction fails as the time gap between the past and the future grows. A change of the observed-to-predicted ratio with the time gap means that the activity of the sources changes in time. The strength of this method is that it needs no assumption on the distribution of burst rates or on the distances of the sources. It needs only the number of times each source is detected, so it applies to any class of repeating transients.

We describe the method in \S\ref{sec:method} and the data in \S\ref{sec:data}. We show in \S\ref{sec:results} that the activity is not steady, compare models of how sources go quiet in \S\ref{sec:models}, and discuss the results in \S\ref{sec:discussion}.

\section{Method} \label{sec:method}

\subsection{Predicting future bursts from past detections}

Consider sources whose detections follow a Poisson process with a constant mean rate $r$. Let $f(r)$ be the number of sources per unit rate. We write $N$ for numbers of sources and $n$ for numbers of bursts. In a window A of exposure $T_A$, the expected number of sources detected exactly $k$ times is
\begin{equation}
N_k = \int f(r)\, \frac{(rT_A)^k}{k!}\, e^{-rT_A}\, dr .
\end{equation}
These sources emit on average $r T_B$ bursts in a second window B of exposure $T_B$. Summing over them gives
\begin{equation}
n^{\rm pred}_k = \int f(r)\, \frac{(rT_A)^k}{k!}\, e^{-rT_A}\, rT_B\, dr = (k+1)\, N_{k+1}\, \frac{T_B}{T_A} . \label{eq:robbins}
\end{equation}
The right-hand side contains only observed counts. Eq.~(\ref{eq:robbins}) is the Robbins formula of unseen species estimation, and it holds for any $f(r)$ \citep{Robbins1956}. It also holds when the exposure differs between sky positions, as long as window B adds exposure in proportion to window A at every position. A continued survey with the same instrument satisfies this condition.

\begin{figure*}
\centering
\includegraphics[width=0.95\textwidth]{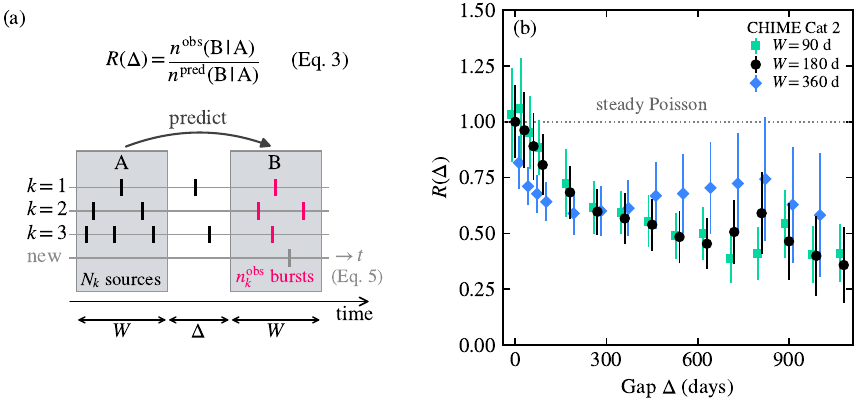}
\caption{(a) Layout of the test. Windows A and B (gray boxes) have length $W$ and are separated by a gap $\Delta$. Each horizontal line is an example source and each tick is a detection. Sources are grouped by their number $k$ of detections in A, and their detections in B (pink ticks) give $n^{\rm obs}_k$. A source first detected in B (gray tick) is a new source and sets the exposure ratio $t$. (b) $R(\Delta)$ of CHIME/FRB Catalog~2 for $W = 90$ days (green squares), 180 days (black circles, our main choice), and 360 days (blue diamonds), shifted slightly in $\Delta$ for clarity. Error bars are bootstrap standard deviations. Gaps with fewer than five window placements are not shown. The dotted line marks $R = 1$, expected for steady Poisson sources.}
\label{fig:concept}
\end{figure*}

\subsection{The gap statistic}

We place windows A and B of equal length $W$ with a gap $\Delta$ between them (Figure~\ref{fig:concept}a). For each placement $s$ we count $N_k(s)$, the number of sources detected $k$ times in window A. We then follow these same sources into window B. We define $n^{\rm obs}_k(s)$ as the total number of bursts that the sources detected $k$ times in A emit in B. For example, a source detected twice in A and three times in B adds three to $n^{\rm obs}_2$. Eq.~(\ref{eq:robbins}) predicts this number. We slide the pair over the survey in steps of 30 days and define
\begin{equation}
R(\Delta) = \frac{n^{\rm obs}}{n^{\rm pred}} = \frac{\sum_s \sum_{k=1}^{3} n^{\rm obs}_k(s)}{\sum_s \sum_{k=1}^{3} t(s)\,(k+1)\, N_{k+1}(s)} , \label{eq:R}
\end{equation}
where $t(s) = T_B/T_A$ is the exposure ratio of placement $s$. The two windows have the same length, but their exposures can differ because the uptime and the sensitivity of CHIME change with time~\citep{CHIME2026cat2}. Therefore, we estimate $t(s)$ from the data (\S\ref{sec:exposure}). We use $k \leq 3$ because sources seen many times in A carry little information on the rate distribution and can dominate the sum. We use $W = 180$ days as a compromise. A shorter window contains too few sources seen two or three times, and in a longer window the activity changes within the window itself (\S\ref{sec:results}). We measure $R$ at $\Delta = 0$, 30, 60, and 90 days and then every 90 days, half a window, up to 1080 days.

Steady Poisson sources give $R = 1$ at every $\Delta$. Burst clustering on timescales longer than a day but shorter than the gap raises $N_2$ and $N_3$ relative to their Poisson values. This lowers $R$ by a similar factor at all gaps longer than the clustering timescale. Sources that stop or switch off remove bursts from window B and make $R$ fall with $\Delta$. Sources born after window A are not counted in $N_k$, so they enter neither $n^{\rm obs}$ nor $n^{\rm pred}$. Therefore, $R$ is not affected by new sources and measures only how long the sources seen in A stay active. We summarize the early decline by the ratio of mean values
\begin{equation}
S = \frac{R(\Delta \geq 360\,\mathrm{d})}{R(\Delta \leq 90\,\mathrm{d})} .
\end{equation}

\subsection{Exposure from the number of new sources} \label{sec:exposure}

The expected number of sources that are not detected in A but are detected in B is
\begin{equation}
N_{\rm new} = \sum_{k \geq 1} (-1)^{k+1}\, t^k N_k , \qquad t = T_B/T_A , \label{eq:et}
\end{equation}
again for any $f(r)$. This is the Good--Toulmin estimator of the number of unseen species \citep{Good1956,Efron1976}. We derive Eq.~(\ref{eq:et}) in Appendix~\ref{app:goodtoulmin}. We solve Eq.~(\ref{eq:et}) for $t(s)$ in each placement, with the series truncated at $k = 8$. The new sources are not detected in A, so the estimate of $t(s)$ does not use $n^{\rm obs}_k$. In our mock catalogs (Appendix~\ref{app:validation}), Eq.~(\ref{eq:et}) predicts $N_{\rm new}$ to within 2\% even with burst clustering or sources that stop.

\section{Data} \label{sec:data}

We use CHIME/FRB Catalog~2 \citep{CHIME2026cat2}, including all bursts from known repeaters. The data span 1837 days from 2018 September 4 to 2023 September 15. We remove bursts with \texttt{excluded\_flag} $= 1$ and keep bursts with S/N $> 10$. This leaves 3714 bursts. We identify a source by its repeater name when one is given and by the burst name otherwise.

CHIME observes each source only during its daily transit through the primary beam, which lasts a few minutes. Active repeaters can emit several bursts within one transit, and the catalog lists the sub-bursts of one event as separate entries. These bursts are not independent, and counting them separately would inflate $N_2$ and $N_3$. Therefore, we merge bursts of one source that arrive within 0.5 days into one detection, so that a detection means a transit with at least one burst and the short-term clustering of bursts within one transit does not enter our counts. The sample has 2686 sources, of which 62 are detected more than once. Merging within 0.02 or 0.3 days instead changes $S$ by at most 0.011.

\section{The activity is not steady} \label{sec:results}

The black circles in Figure~\ref{fig:concept}b show $R(\Delta)$ for $W = 180$ days. $R$ is consistent with unity for $\Delta \leq 60$ days, falls to $0.57 \pm 0.11$ at $\Delta = 360$ days, and stays near one half out to $\Delta = 900$ days. At the longest gap of 1080 days it is $0.36 \pm 0.17$. The ratio is $S = 0.530 \pm 0.142$. The uncertainty is the standard deviation of 500 bootstrap samples drawn by resampling sources, and a jackknife over sources gives 0.147. The estimated exposure ratios $t(s)$ average between 0.93 and 1.04 at all gaps. A change of sensitivity during the survey would also change the number of new sources, so $t(s)$ absorbs it in each placement (\S\ref{sec:exposure}). Therefore, neither the decline nor the flattening of $R$ beyond about one year comes from instrumental effects.

The decline does not depend on the choice of $W$ (Figure~\ref{fig:concept}b). With $W = 90$ days, $R$ follows the values for $W = 180$ days within the errors, and $S = 0.49 \pm 0.12$. With $W = 360$ days, $R$ is already $0.82 \pm 0.12$ at $\Delta = 0$, because the activity of many sources changes within one window. At long gaps the errors are large, because only a few placements of such long windows fit in the survey. Possible missed associations in the catalog, which turn a repeat into two one-off events, could also lower $R$ at long gaps. We searched for them in position and dispersion measure (DM), including the candidate repeaters of \citet{Cook2026}, and found only a few beyond chance. Treating all candidates as repeaters changes $S$ by at most 0.05 (Appendix~\ref{app:assoc}).

We use the ratio $S$ to test whether the activity is steady, because burst clustering changes the level of $R$ but hardly its shape (\S\ref{sec:method}), so $S$ tests for a change of activity without a model of the clustering. To assess the significance, we compare $S$ with mock catalogs of steady sources. The burst rates follow a power law $f(r) \propto r^{s}$ between $10^{-6}$\,day$^{-1}$ and $r_{\max}$. Here $r$ is the rate of detected bursts, which combines the intrinsic activity, the energy distribution, the distance, and the exposure of each source. We choose $s$ and $r_{\max}$ on a grid, and the number of sources, so that the mocks reproduce the mean counts $N_1$, $N_2$, $N_3$, and $N_{\geq 4}$ per 180-day window of the data. The fits give $s \approx -2.3$. Eq.~(\ref{eq:robbins}) does not depend on $f(r)$, so this choice does not enter the measured $R(\Delta)$ or the expectation $R = 1$. It affects only the scatter of $S$ in the mocks and hence the $p$-values.

\begin{figure}
\centering
\includegraphics[width=0.46\textwidth]{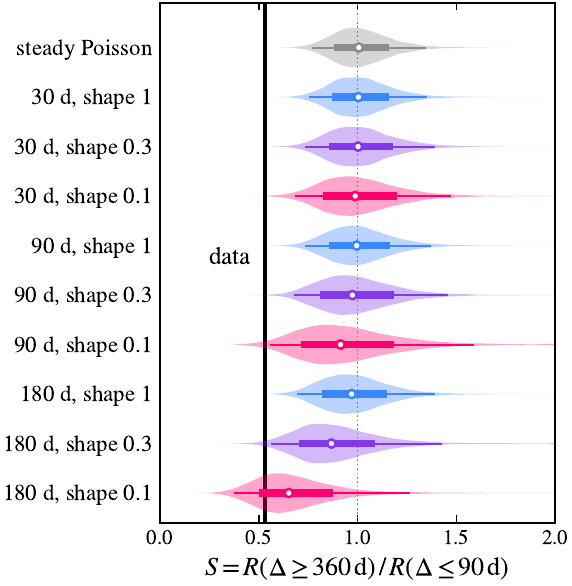}
\caption{Distribution of the ratio $S$ in 10,000 mock catalogs of sources that never stop, for each model of emission. Steady Poisson means a constant burst rate. In the other models the rate in each epoch of the given length is multiplied by a random factor from a Gamma distribution with the given shape. A smaller shape means stronger variation between epochs (blue for 1, purple for 0.3, pink for 0.1). Shaded areas show the full distribution, thick bars the 16th to 84th percentiles, thin bars the 2.5th to 97.5th percentiles, and white circles the medians. The black vertical line marks the data.}
\label{fig:null}
\end{figure}

Figure~\ref{fig:null} shows the results. For steady Poisson sources, no mock catalog out of 10,000 has $S \leq 0.530$ ($p < 10^{-4}$). Repeaters cluster their bursts in time \citep{Yamasaki2026sc}, so we also test episodic emission, in which the rate in each epoch of fixed length is multiplied by a random factor from a Gamma distribution with mean 1. For epochs of 30, 90, and 180 days and Gamma shapes of 1.0, 0.3, and 0.1, all models give $p \leq 0.015$ except 180-day epochs with a shape of 0.1 ($p = 0.22$). In that model the burst rate is below a tenth of its mean in about two thirds of the epochs, and the forecast test of \S\ref{sec:models} does not reject it either ($p = 0.2$), while it rejects 180-day epochs with a shape of 0.3 and 90-day epochs with a shape of 0.1 ($p \leq 0.02$). Such sources are silent most of the time, which is itself a form of the on/off activity of \S\ref{sec:models}. Therefore, the data require that most sources are quiet most of the time, whether they stop for good or switch off and on.

\section{How do sources go quiet?} \label{sec:models}

We compare three families of models in which sources go quiet.
\begin{enumerate}
\item \emph{Single lifetime.} Sources switch on at a constant rate and stop for good after an exponentially distributed lifetime with mean $\tau$.
\item \emph{On/off activity.} Each source alternates between active phases with an exponentially distributed length of mean $T_{\rm on}$ and quiet phases of mean $T_{\rm off} = T_{\rm on}(1-d)/d$, where $d$ is the duty cycle. While active, the source bursts at rate $r/d$, so its long-term mean rate is $r$.
\item \emph{Mixture of lifetimes.} Of the sources active at any time, a fraction $\eta$ keeps going through the survey. The rest stop after an exponentially distributed lifetime with mean $\tau_s$ and are replaced by new sources.
\end{enumerate}
In all models the bursts of an active source follow a Poisson process, and the number of active sources is constant on average, because new sources switch on at the rate at which others stop. The survey of 5\,yr is very short compared with the cosmic time over which the FRB population changes, so this is a natural assumption. It enters only the mock catalogs, not the measured $R(\Delta)$ (\S\ref{sec:method}). We compute each model on a grid of its parameters. At each grid point we tune $f(r)$ to the window counts as before and measure $R(\Delta)$ in 100 to 200 mock catalogs with the same code as for the data.

We compare each model with the data through a $\chi^2$ that uses the bootstrap covariance of $R(\Delta)$. The errors at different gaps are strongly correlated, and the small eigenvalues of the covariance are poorly measured. Therefore, we keep only the three leading eigenmodes of the covariance, which describe the level, the slope, and the curvature of $R(\Delta)$. In mock tests this gives unbiased and conservative parameter ranges (Appendices~\ref{app:chi2} and \ref{app:validation}).

\begin{figure*}
\centering
\includegraphics[width=\textwidth]{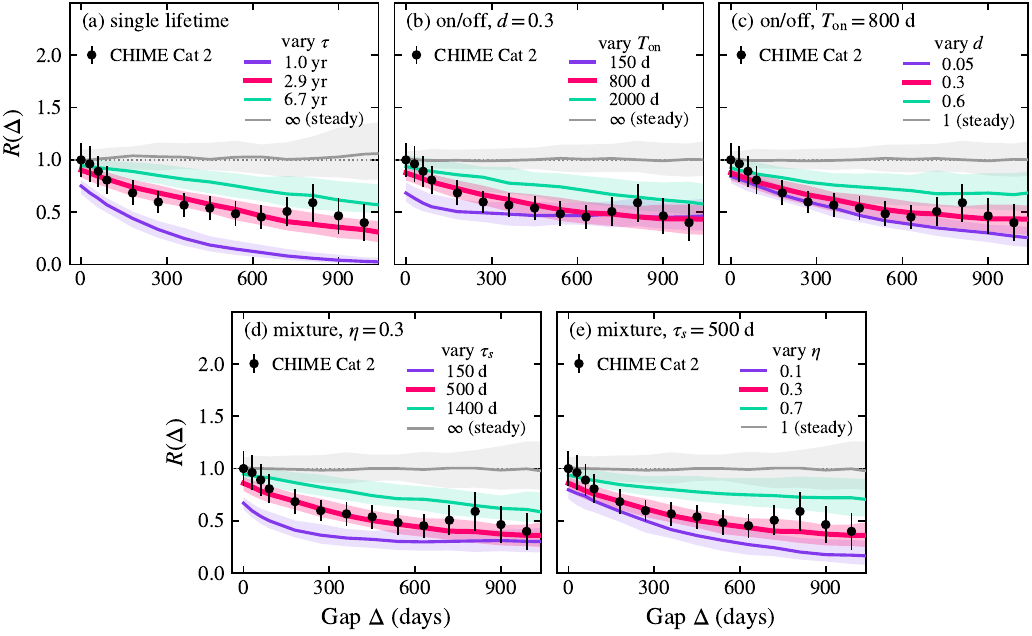}
\caption{Black circles are $R(\Delta)$ of CHIME/FRB Catalog~2 for $W = 180$ days, as in Figure~\ref{fig:concept}b. Lines show the median of mock catalogs, and the shaded band of the same color shows their 16th to 84th percentiles. In each panel one parameter of a model is varied and the other is fixed at its best value, given in the panel title. The thick pink line is the best model of that family. Other values go from purple (small) through blue to green (large), as listed in the legend. Gray lines are steady sources, which give $R = 1$ (dotted line). (a) A single activity lifetime $\tau$. (b) On/off activity with the mean active phase $T_{\rm on}$ varied. (c) On/off activity with the duty cycle $d$ varied. (d) A mixture of short-lived and long-lived sources with the mean lifetime $\tau_s$ of the short-lived sources varied. (e) The same mixture with the long-lived fraction $\eta$ varied.}
\label{fig:models}
\end{figure*}

The thick pink lines in Figure~\ref{fig:models} show the best model of each family. All three fit the data, with $\chi^2 = 0.4$ for on/off activity, 0.5 for the mixture, and 1.5 for the single lifetime, for three modes. The differences are too small to prefer one model. The steady model gives $\chi^2 = 16$ to 18 and is ruled out, in agreement with \S\ref{sec:results}. The best models of each family also fit the data for $W = 360$ days, with $\chi^2 = 2.7$ to 3.9 for three modes, while the steady model gives $\chi^2 = 18$ to 21. The longer window confirms the decline but does not separate the three models.

Figure~\ref{fig:models} shows how each parameter changes $R(\Delta)$. The timescale sets where $R$ starts to fall (panels a, b, and d). Only the middle values follow the data at both short and long gaps. The second parameter sets where $R$ levels off (panels c and e). In the on/off model, a source seen in window A is in an active phase. At a much later time it is active again with probability $d$, independent of its state in A, so $R$ approaches $d$ at gaps much longer than $T_{\rm on}$. In the mixture, only the long-lived sources still burst at long gaps. They make up a fraction $\eta$ of the active sources and have the same rate distribution as the others, so $R$ approaches $\eta$. Both limits hold when $W$ is short compared with $T_{\rm on}$ and $\tau_s$, and the mock catalogs of a longer survey follow them (\S\ref{sec:discussion}). If future data show $R$ leveling off, its level gives the duty cycle or the long-lived fraction directly.

We also tested whether the models predict data that they have not seen. We fitted them only to placements with both windows in the first 1100 days (open circles in Figure~\ref{fig:forecast}a), which give $R$ at gaps up to 540 days. We then compared their prediction with $R$ from placements with window A in the first 1100 days and window B in the remaining 738 days (filled circles), which the fit does not use. The models fitted to the early data predict how $R$ keeps falling out to gaps of 1080 days, well beyond the gaps used in the fit. The early data alone do not rule out steady sources, but steady sources predict no such fall. To measure this without the overall level of $R$, we divide the mean $R$ of the later data at $\Delta \geq 630$ days by that of the early data at $\Delta \leq 90$ days. This ratio is 0.41 for the data, 0.41 to 0.59 for the three models, and 1.01 (0.85 to 1.20) for steady sources. No mock catalog of steady sources out of 1000 reaches the data ($p < 10^{-3}$), and splitting at 900 or 1200 days gives $p \leq 0.002$.

The three models agree on the timescale. With 68\% ranges of the posterior, the single lifetime gives $\tau = 2.9^{+1.1}_{-0.9}$\,yr. The on/off model gives $T_{\rm on} = 800^{+400}_{-400}$ days, while the duty cycle is only weakly constrained ($d = 0.3^{+0.05}_{-0.2}$). The mixture gives $\tau_s = 500^{+300}_{-200}$ days and $\eta = 0.3^{+0.2}_{-0.1}$. Therefore, the activity of a typical repeater declines on a timescale of one to three years. The parameter values depend on how the models are compared with the data (Appendix~\ref{app:models}), so we regard them as indicative.

\section{Discussion} \label{sec:discussion}

The three models describe different physics. A single lifetime means that each source has one active stage of a few years and then stops for good. A few years is much shorter than the ages of Galactic magnetars \citep{Beniamini2019}, so this stage is more likely one episode of strong activity than the whole life of the source. Outbursts of Galactic magnetars decay over months to years \citep{KaspiBeloborodov2017,CotiZelati2018}, and SGR~1935+2154 emitted an FRB-like burst during such an active episode \citep{CHIME2020sgr,Bochenek2020}. Therefore, repeaters may be magnetars caught in a short active phase \citep[e.g.,][]{Margalit2020,Lu2020}. In the toy model of \citet{Popov2023}, the most active repeaters are magnetars in a few-year active phase within a much longer lifetime of one-off bursting. Our timescale is comparable, although most repeaters in our sample are less active. A very young magnetar whose activity fades as it ages is another possibility \citep{Murase2016,Beloborodov2017,Kashiyama2017,Metzger2017,Nicholl2017,Margalit2018,BeniaminiKumar2025}. A decrease of the DM seen in several repeaters may also point to young sources in expanding supernova remnants \citep{Cui2026}. On/off activity means that sources return after quiet phases of years. A mixture means that a minority of sources stays active much longer than the rest, as would happen for a spread of ages or of energy reservoirs.

Other analyses of CHIME/FRB data also point to a finite activity of repeaters. \citet{Cook2026} detected no new bursts from 41\% of the previously known CHIME repeaters in the later part of the survey and pointed to their apparent finite lifetimes. Recently, \citet{Yamasaki2026sc} obtained an independent result from the same catalog. They analyzed the interval between the first two detected bursts of each repeater with an on/off model in which bursts are clustered within active phases. They found $T_{\rm on} = 400^{+190}_{-70}$ days and $d \lesssim 0.07$, with $d$ up to about 0.2 allowed at 95\%. Their $T_{\rm on}$ is about half our best value of 800 days but agrees with it within the errors. The two statistics are sensitive to different times. The interval between the first two bursts is set mostly within the first active phase that is seen, while $R(\Delta)$ follows the same sources over several years. The models also differ, because bursts in our model are not clustered within active phases. Our own $T_{\rm on}$ ranges from about 300 to 800 days with the statistic used (Appendix~\ref{app:models}), so a factor of two is within the systematic uncertainty. The duty cycles overlap near 0.1 to 0.2.

The timescale we measure should not be confused with the periodic activity of some repeaters. FRB~20180916B and FRB~20121102A are active in windows that recur every 16.3 and about 160 days \citep{CHIME2020periodic,Rajwade2020}. For the hyperactive FRB~20240114A, candidate periods of about 70 to 140 days have been reported in the activity, the emission frequency, and the degree of linear polarization \citep{Zhou2026,Li2026freq,Sun2026}. These periods are often attributed to the geometry of the source, such as the orbit of a binary \citep{IokaZhang2020,Lyutikov2020,Wada2021} or the precession of a magnetar \citep{Levin2020,ZanazziLai2020}. Our windows of 180 days are as long as or longer than these periods and average over them. The change of activity over one to three years that we find is therefore a different effect. It more likely reflects a change of the engine itself, such as the evolving activity of a magnetar. The two effects can coexist in one source, with periodic windows inside a longer active phase.

The models also imply different numbers of sources. The number density of active sources is $n_{\rm src} = \mathcal{R}_{\rm FRB}/(f_{\rm b}\,\bar{r})$, where $\mathcal{R}_{\rm FRB} = 7.3 \times 10^{4}$\,Gpc$^{-3}$\,yr$^{-1}$ is the volumetric rate of bursts above $10^{39}$\,erg from CHIME/FRB \citep[][see also \citealt{Hashimoto2022}]{Shin2023}, $f_{\rm b}$ is the beaming fraction, and $\bar{r}$ is the mean rate of such bursts per source. We compare the rate $\mathcal{R}_{\rm birth}$ at which new sources must switch on with the birth rate of magnetars, $\mathcal{R}_{\rm mag} = f_{\rm mag}\, \mathcal{R}_{\rm CCSN}$. Here $\mathcal{R}_{\rm CCSN} \approx 8 \times 10^{4}$\,Gpc$^{-3}$\,yr$^{-1}$ is the local rate of core-collapse supernovae \citep{Frohmaier2021,Ma2025}, and $f_{\rm mag}$ is the fraction of neutron stars born as magnetars, estimated between 0.12 and nearly 1 \citep{Beniamini2019,PardoAraujo2026}, and we assume that each core-collapse supernova leaves a neutron star. For a single lifetime, $\mathcal{R}_{\rm birth} = n_{\rm src}/\tau$, and therefore
\begin{equation}
\frac{\mathcal{R}_{\rm birth}}{\mathcal{R}_{\rm mag}} \sim 0.1 \left(\frac{\mathcal{R}_{\rm FRB}}{\mathcal{R}_{\rm CCSN}}\right) \left(\frac{0.1}{f_{\rm b}}\right) \left(\frac{0.4}{f_{\rm mag}}\right) \left(\frac{100\,{\rm yr}^{-1}}{\bar{r}}\right) ,
\end{equation}
where we used $\tau \approx 2.9$\,yr. The volumetric rates of FRBs and core-collapse supernovae are similar, so the ratio is set by $f_{\rm b} f_{\rm mag} \bar{r} \tau$.
For the mixture, the short-lived sources set the birth rate, $\mathcal{R}_{\rm birth} \approx (1-\eta)\, n_{\rm src}/\tau_s$, which gives a similar ratio for $\tau_s \approx 500$ days and $\eta \approx 0.3$. For plausible $f_{\rm b}$ and $\bar{r}$, only a minority of newborn magnetars needs to pass through an active stage of a few years, but the large uncertainties of these values allow a ratio near unity. For on/off activity, the sources, active or quiet, are $1/d$ times, or three to ten times, as many as the active ones. Their birth rate depends on their total lifetime, which our data do not constrain.

\begin{figure*}
\centering
\includegraphics[width=0.95\textwidth]{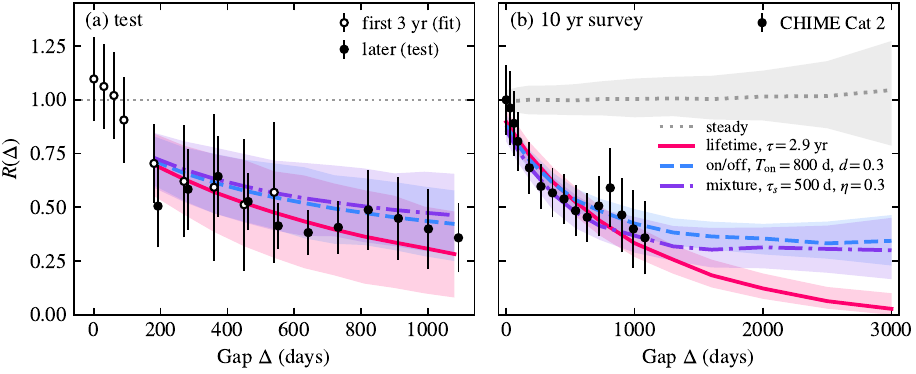}
\caption{Predictions of $R(\Delta)$ with $W = 180$ days. Lines show the median of mock catalogs and shaded bands show their 16th to 84th percentiles, for a single lifetime (pink solid), on/off activity (blue dashed), and a mixture of short-lived and long-lived sources (purple dash-dotted). (a) Test within Catalog~2. Open circles are $R$ from the first 1100 days, to which the models are fitted. Filled circles are $R$ with window B in the remaining 738 days, which the fit does not use. Lines and bands show the prediction weighted by the posterior of the fit. The gray dotted line marks $R = 1$, expected for steady sources. (b) Forecast for a survey of 10\,yr with the detection rate of Catalog~2, for the best parameters found from Catalog~2. The gray dotted line and band show steady sources. Black circles are the data of Catalog~2.}
\label{fig:forecast}
\end{figure*}

Our data cannot separate the three models, but longer and more sensitive surveys can. We drew mock catalogs of the best model of each family for a survey of 10\,yr with the same window counts as Catalog~2 (Figure~\ref{fig:forecast}b). For the single lifetime, $R$ keeps falling and drops below 0.1 beyond about 2500 days. For on/off activity and the mixture, it levels off near 0.3. With 10\,yr of CHIME/FRB data, the single lifetime differs from on/off activity by about $3\sigma$ and from the mixture by about $1.6\sigma$. If each source is detected three times as often, these differences grow to about $5\sigma$ and $3.4\sigma$. Such a gain can come from a more sensitive telescope such as CHORD \citep{Vanderlinde2019} or from a wide-field telescope that watches each source for longer every day such as BURSTT \citep{Lin2022,Wang2026,Lin2026}, DSA/Chronoscope
\citep{connor2026}, and CHARTS \citep{charts1,charts2}.
On/off activity and the mixture give almost the same $R$ (within $0.6\sigma$), so $R$ alone cannot separate them. On/off activity also predicts that repeaters silent for several years will reappear at their old positions and DMs, while the other two models predict that they will not. Future CHIME/FRB catalogs can test this by searching for repeaters of Catalog~1 and Catalog~2 that return after long silence.

Our measurement concerns only sources that repeat. Truly one-off events, if they exist, add only to $N_1$ and enter neither $n^{\rm obs}$ nor $n^{\rm pred}$. They also do not bias $t$, because Eq.~(\ref{eq:et}) holds for them as for sources with very low rates. A source that is active for a year or two and then quiet will often be detected once and then not again for years. Therefore, some one-off sources can be repeaters that have gone quiet, as suggested by population studies in which most or all CHIME sources can be repeaters \citep{James2023,Yamasaki2024,BeniaminiKumar2025,Cook2026,Yamasaki2026sc}. It also implies that fewer one-off sources will repeat in the coming years than a steady model predicts. Future catalogs can test this prediction with Eq.~(\ref{eq:robbins}) and the same windows.

\section{Conclusions} \label{sec:conclusions}

We tested whether repeating FRB sources keep a steady activity, using unseen species estimation on CHIME/FRB Catalog~2. The test uses only the numbers of sources detected $k$ times and needs no assumption on the distribution of burst rates or on the distances of the sources. The ratio of observed to predicted bursts is close to unity at gaps of $\lesssim 60$ days and falls to about one half within one year. This rules out steady Poisson sources ($p < 10^{-4}$). Sources that never stop but whose rate changes randomly every 30 to 180 days are also ruled out ($p \leq 0.015$), unless they are silent most of the time. A lifetime of about 3\,yr, on/off activity with active phases of about two years, and a mixture of short-lived and long-lived sources all fit the data. Models fitted to the first three years predict the decline in the later data. Therefore, the activity of repeaters declines on a timescale of one to three years. This timescale is consistent with a short active phase of magnetars, which their birth rate can supply. It also implies that some apparently one-off sources are repeaters that have gone quiet. Longer surveys will show whether quiet sources stop for good or become active again.

Our results also suggest when to observe repeaters. The expected number of bursts falls by about half within a year, so repeaters should be followed up soon after they are found. Sources that have not burst for several years are less promising targets, unless the goal is to catch a source that returns after a long silence. Unlike the butterflies of \citet{Fisher1943}, whose species were still there on the next collecting trip, many FRB sources seen today may have gone quiet by the time we return. For repeaters, the next collecting trip should not wait long.

\begin{acknowledgments}
We thank the CHIME/FRB Collaboration for making Catalog~2 public. SY acknowledges the support from the National Science and Technology Council of Taiwan (NSTC) through grant numbers 113-2112-M-005-007-MY3 and 113-2811-M-005-006-.
\end{acknowledgments}

\appendix

\section{Derivation of the Good--Toulmin estimator} \label{app:goodtoulmin}

Consider a source with a constant rate $r$ and windows A and B with exposures $T_A$ and $T_B = t\,T_A$. For a Poisson source, the probability that it is not detected in A, $e^{-rT_A}$, and the probability that it is detected in B, $1 - e^{-rT_B}$, are independent. Therefore, the expected number of new sources is
\begin{equation}
N_{\rm new} = \int f(r)\, e^{-rT_A} \left(1 - e^{-r t T_A}\right) dr .
\end{equation}
Expanding $1 - e^{-x} = \sum_{k \geq 1} (-1)^{k+1} x^k / k!$ with $x = r t T_A$ gives
\begin{equation}
N_{\rm new} = \sum_{k \geq 1} (-1)^{k+1}\, t^k \int f(r)\, \frac{(rT_A)^k}{k!}\, e^{-rT_A}\, dr = \sum_{k \geq 1} (-1)^{k+1}\, t^k N_k ,
\end{equation}
where we used the definition of $N_k$ in \S\ref{sec:method}. No assumption on $f(r)$ enters. The exchange of the sum and the integral is valid for $t \leq 1$, where the series converges. For $t > 1$ the series can diverge \citep{Good1956,Efron1976}. In our analysis $t$ is close to one, and the terms with large $k$ are small, because few sources are detected many times in one window.

\section{Comparing models with correlated errors} \label{app:chi2}

We compare each model with the data through
\begin{equation}
\chi^2 = (\mathbf{R} - \mathbf{M})^{\rm T}\, \mathbf{C}^{-1}\, (\mathbf{R} - \mathbf{M}) , \label{eq:chi2}
\end{equation}
where $\mathbf{R}$ is the vector of the data at the 15 gaps, $\mathbf{M}$ is the vector of the median of the mock catalogs, and $\mathbf{C}$ is the covariance matrix of $\mathbf{R}$ from 500 bootstrap samples of the data. Because $R$ at neighboring gaps comes from almost the same placements, $\mathbf{C}$ has very small eigenvalues for the differences between neighboring gaps. These modes are dominated by noise, and a full inverse of $\mathbf{C}$ lets them control $\chi^2$. Therefore, we keep only the three leading eigenmodes of $\mathbf{C}$ and replace $\mathbf{C}^{-1}$ in Eq.~(\ref{eq:chi2}) by
\begin{equation}
\mathbf{C}^{-1} \simeq \sum_{m=1}^{3} \frac{\mathbf{v}_m \mathbf{v}_m^{\rm T}}{\lambda_m} ,
\end{equation}
where $\lambda_m$ are the three largest eigenvalues and $\mathbf{v}_m$ are their eigenvectors. These three modes carry 94\% of the variance (62\%, 25\%, and 6\%). Their eigenvectors are close to a constant, a straight line, and a parabola in $\Delta$, with normalized overlaps of 0.98, 0.90, and 0.90, so they describe the level, the slope, and the curvature of $R(\Delta)$.

\section{Possible missed associations} \label{app:assoc}

The catalog may list two bursts of one source as different sources, if their positions or DMs were not matched. Such missed associations turn a repeat into two one-off events and, if they were more common at long separations, would lower $R$ at long gaps. We call two bursts a candidate match if their positions agree within $0.5^\circ$ or within three times their combined position uncertainty, whichever is larger, and their DMs differ by less than 5\,pc\,cm$^{-3}$. We estimate the number of chance matches from control samples in which the DM of one burst is shifted by $\pm 25$ and $\pm 40$\,pc\,cm$^{-3}$.

Eight one-off bursts match a known repeater, compared with 2.2 expected by chance. Eighteen pairs of one-off bursts separated by more than 0.5 days match each other, compared with 13.9 expected by chance, with no trend in their separation. Assigning the 8 bursts to their repeaters gives $S = 0.523$, and merging all candidate pairs into common sources, a worst case that includes the chance matches, gives $S = 0.577$. Of the 14 candidate repeaters of \citet{Cook2026}, five already carry a repeater name in the catalog. Treating them as one-off sources leaves $S = 0.530$. Among the other nine, two keep two detections after the S/N cut, and treating them as repeaters gives $S = 0.534$.

\section{Validation and model details} \label{app:validation} \label{app:models}

In 300 mock catalogs of steady Poisson sources, the normalized residuals of Eqs.~(\ref{eq:robbins}) and (\ref{eq:et}) have a mean within 0.1 of zero and a standard deviation of 0.9 to 1.07. For sources that stop or cluster their bursts, Eq.~(\ref{eq:et}) still predicts the number of new sources to within 2\%. To test the model comparison, we fitted each of the 100 mock catalogs of an on/off grid point ($T_{\rm on} = 400$ or 800 days) on the whole grid. The median fit recovers the true values within one grid step. With the three leading eigenmodes, the 68\% range of $T_{\rm on}$ contains the true value in 82\% to 94\% of the mocks, while with the full inverse covariance it does so in only 50\% to 70\%.

\begin{figure*}
\centering
\includegraphics[width=0.9\textwidth]{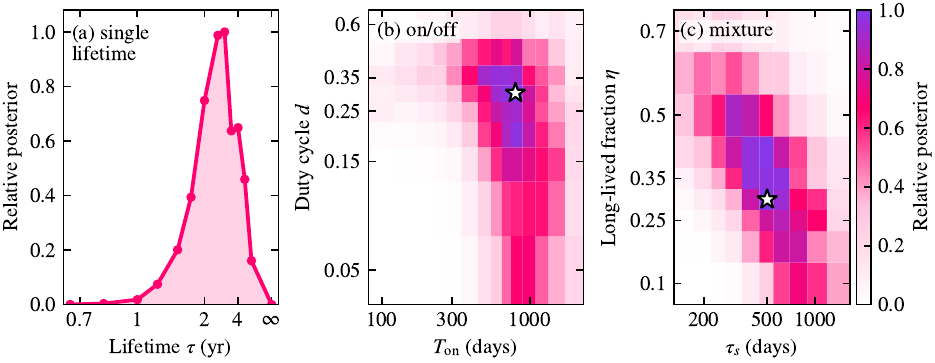}
\caption{Posteriors of the three model families, $\propto \exp(-\chi^2/2)$ with a flat prior over the grid and the three leading eigenmodes (Appendix~\ref{app:chi2}). (a) Single lifetime $\tau$, on an axis uniform in $1/\tau$ (the right end is the steady model). (b) On/off activity in $T_{\rm on}$ and $d$. (c) Mixture in $\tau_s$ and $\eta$. In (b) and (c) the color gives the posterior relative to its maximum and the star marks the best grid point.}
\label{fig:posteriors}
\end{figure*}

Figure~\ref{fig:posteriors} shows the posteriors of the three model families. $T_{\rm on}$ and $d$ are partly degenerate. The parameters depend on the statistic. For the on/off model, ratios of $R$ that ignore its level prefer $T_{\rm on} \approx 600$ days, and the $\chi^2$ with the full inverse covariance prefers about 300 days with too narrow ranges, as the mock tests above show. All statistics rule out steady sources. A broken power law $f(r)$ gives nearly the same results as a single power law, while a log-normal $f(r)$ cannot reproduce the window counts.

%\bibliography{main}{}
\bibliographystyle{aasjournal}

\end{document}